\documentclass[preprint1]{aastex}
\usepackage{amsmath}
\slugcomment{}

\newcommand{\be}{\begin{equation}}
\newcommand{\ee}{\end{equation}}

\shorttitle{Type Ic Super Luminous Supernovae}
\shortauthors{Wei, Wu \& Melia}

\begin{document}

\title{Testing Cosmological Models with Type Ic Super Luminous Supernovae}
\author{Jun-Jie Wei\altaffilmark{1,3}, Xue-Feng Wu\altaffilmark{1,4,5} and
Fulvio Melia\altaffilmark{1,2}}
\altaffiltext{1}{Purple Mountain Observatory, Chinese Academy of Sciences, Nanjing 210008, China.}
\altaffiltext{2}{Department of Physics, The Applied Mathematics Program, and Department of Astronomy,
The University of Arizona, AZ 85721, USA; fmelia@email.arizona.edu.}
\altaffiltext{3}{University of Chinese Academy of Sciences, Beijing 100049, China; jjwei@pmo.ac.cn.}
\altaffiltext{4}{Chinese Center for Antarctic Astronomy, Nanjing 210008, China; xfwu@pmo.ac.cn.}
\altaffiltext{5}{Joint Center for Particle, Nuclear Physics and Cosmology, Nanjing
University--Purple Mountain Observatory, Nanjing 210008, China.}

\begin{abstract}
The use of type Ic Super Luminous Supernovae (SLSN Ic) to examine the cosmological
expansion introduces a new standard ruler with which to test theoretical
models. The sample suitable for this kind of work now includes 11 SLSNe Ic,
which have thus far been used solely in tests involving $\Lambda$CDM.
In this paper, we broaden the base of support for this new, important cosmic
probe by using these observations to carry out a one-on-one comparison
between the $R_{\rm h}=ct$ and $\Lambda$CDM cosmologies. We individually
optimize the parameters in each cosmological model by minimizing the $\chi^{2}$
statistic. We also carry out Monte Carlo simulations based on these current
SLSN Ic measurements to estimate how large the sample would have to be in
order to rule out either model at a $\sim 99.7\%$ confidence level. The currently
available sample indicates a likelihood of $\sim$$70-80\%$ that the
$R_{\rm h}=ct$ Universe is the correct cosmology versus $\sim$$20-30\%$ for the
standard model. These results are suggestive, though not yet compelling, given the
current limited number of SLSNe Ic. We find that if the real cosmology is $\Lambda$CDM,
a sample of $\sim$$240$ SLSNe Ic would be sufficient to rule out $R_{\rm h}=ct$ at
this level of confidence, while $\sim$$480$ SLSNe Ic would be required to rule out
$\Lambda$CDM if the real Universe is instead $R_{\rm h}=ct$. This difference
in required sample size reflects the greater number of free parameters available
to fit the data with $\Lambda$CDM. If such SLSNe Ic are commonly detected in the
future, they could be a powerful tool for constraining the dark-energy equation of state
in $\Lambda$CDM, and differentiating between this model and the $R_{\rm h}=ct$
Universe.
\end{abstract}

\section{Introduction}
A new type of luminous transient has been identified in recent years through the use
of deep, wide surveys searching for supernovae in the local Universe (Quimby et al.
2005, 2007; Quimby 2006; Smith et al. 2007; Drake et al. 2009; Rau et al. 2009; Kaiser et
al. 2010; Gal-Yam 2012; Baltay et al. 2013). These super-luminous supernovae (SLSN)
have peak magnitudes $M_{AB}<-21$ mag and integrated burst energies
$\sim$$10^{51}$ erg. They are therefore much brighter than both Type Ia SNe and
the majority of core-collapse events. Their unusually high peak luminosities,
hot blackbody temperatures and bright rest frame ultra-violet emission (which
renders their continuum easily detectable at optical and near-infrared wavelengths
at high redshifts; Cooke et al. 2012) allow them to be studied in concert
with possible gamma-ray burst associations (Li et al. 2014; Cano \& Jakobsson
2014) and, more importantly, make them viable standardizable candles
and distance indicators for use as cosmological probes (Inserra \& Smartt 2014).

Inserra et al. (2013) and Nicholl et al. (2013) used the classification term SLSN Ic
to refer to all the hydrogen poor SNLSe, though there appear to be at least two
observational groups in this category. These are distinguished via the terms
SN2005ap-like and SN2007bi-like events, since these are the prototypes with
the faster and slower evolving lightcurves. SLSNe Ic have now been discovered
out to redshifts $z\sim$$4$ (Chomiuk et al. 2011; Berger et al. 2012;
Cooke et al. 2012; Howell et al. 2013), and appear to be rather homogeneous
in their spectroscopic and photometric properties.

The brighter events decline more slowly, not unlike the Phillips relation
for Type Ia SNe, which raises the possibility of correlating their peak magnitudes
to their decline over a fixed number of days to reduce the scatter (Rust 1974;
Pskovskii 1977; Phillips 1993; Hamuy et al. 1996). Without the use of such a
relation, the uncorrected raw mean magnitudes show a dispersion of $\sim$$0.4$
(e.g., Inserra \& Smartt 2014).  Correlating the peak magnitude to the
decline over 30 days reduces the scatter in standardized peak magnitudes
to $\pm0.22$ mag. And apparently using a magnitude-color evolution reduces
this scatter even more, to a low value somewhere between $\pm0.08$ mag
and $\pm0.13$ mag.

It is therefore quite evident that SLSNe Ic may be useful cosmological probes,
perhaps even out to redshifts much greater ($z>>2$) than those accessible
using Type Ia SNe. The currently available sample, however, is still quite
small; adequate data to extract correlations between empirical, observable
quantities, such as lightcurve shape, color evolution and peak luminosity,
are available only for tens of events. Our focus in this paper is specifically
to study whether SLSNe Ic can be used---not only to optimize the parameters
in $\Lambda$CDM (Li et al. 2014; Cano \& Jakobsson 2014; Inserra \&
Smartt 2014), e.g., to refine the dark-energy equation of state but,
also---to carry out comparative studies between competing cosmologies,
such as $\Lambda$CDM versus the $R_{\rm h}=ct$ Universe (Melia 2007;
Melia \& Abdelqader 2009; Melia \& Shevchuk 2012; Melia 2013a).

Like $\Lambda$CDM, the $R_{\rm h}=ct$ Universe is a Friedmann-Robertson-Walker
cosmology that assumes the presence of dark energy, as well as matter and
radiation. The principle difference between them is that the latter is also
constrained by the equation of state $\rho+3p=0$, in terms of the total pressure
$p$ and energy density $\rho$. In recent years, this model has generated some
discussion concerning its fundamental basis, including claims that it is actually
a vacuum solution, even though $\rho\not=0$. However, all criticisms leveled
against $R_{\rm h}=ct$ thus far appear to be based on incorrect assumptions
and theoretical errors. A full accounting of this discussion may be found
in Melia (2015) and references cited therein.

In fact, the application of model selection tools in one-on-one comparisons
between these two cosmologies has shown that the data tend to favor $R_{\rm h}=ct$
over $\Lambda$CDM. Tests completed thus far include high-$z$ quasars (Melia
2013b, 2014), gamma ray bursts (Wei et al. 2013), the use of cosmic
chronometers (Melia \& Maier 2013) and, most recently, the Type Ia SNe
themselves (Wei et al. 2014). In all of these tests, information
criteria show that $R_{\rm h}=ct$, with the important additional
constraint $\rho+3p=0$ on its equation of state, is favored over
$\Lambda$CDM with a likelihood of $\sim 90\%$ versus only $\sim 10\%$.

Here, we broaden the comparison between $R_{\rm h}=ct$ and
$\Lambda$CDM by now including SLSNe Ic in this study. In \S~2 we
briefly describe the currently available sample and our method
of analysis. Our results are presented in \S~3. We will find that
the current catalog of SLSNe Ic suitable for this study already
confirms the tendencies discussed above, though the statistics
are not yet good enough to strongly differentiate between these
two competing models. We show in \S~4 how large the source
catalog needs to be  in order to rule out one or the other expansion
scenario at a 3-sigma confidence level, and we present our conclusions
\break\noindent\null \null \null in \S~5.

\section{Methodology}
Eleven of the SLSNe Ic identified by Inserra \& Smartt (2014) are appropriate
for this work, and we base our analysis on the methodology described in their
paper and in Li et al. (2014). Briefly, the chosen SLSNe Ic must have well
sampled light-curves around peak luminosity, and photometric coverage from
several days pre-maximum to 30 days (in the rest frame) after the peak.
This time delay appears to be optimal for use in the Phillips-like
peak magnitude-decline relation (Inserra \& Smartt 2014).

All 11 of these events appear to be similar to the well-observed SN2010gx,
and these decay rapidly after peak brightness. They belong to the group
of 2005ap-like events, the first such SN discovered in this category. Other
SLSNe Ic could not be included simply because of lack of sufficient
temporal coverage, even though their identification has been securely
classified in previous work (see, e.g., Leloudas et al. 2012; Quimby et al
2011; Inserra et al. 2013; Chomiuk et al. 2011; Berger et al. 2012;
Nicholl et al. 2014).

\begin{deluxetable}{lcccccccccc}
\tablewidth{514pt}
\tabletypesize{\scriptsize}
\tablecaption{Sample of SLSNe Ic}\tablenum{1}
\tablehead{SN& $z$ & $E(B-V)$  & $m_f$ & Filter & $A_{f}$ & $K^{\rm peak}_{f\rightarrow 400}$ & $m(400)$ & $\Delta M_{30}(400)$ &
$\Delta M_{30}(400-520)$ & Refs.}
\startdata
&&&&&&&&&& \\
{\rm SN}2011{\rm ke} & 0.143 & 0.01 & 17.70 (g) & $g\rightarrow 400$ & 0.05 & -0.18 & 17.83$\pm$0.20 & 2.47$\pm$0.16 & 0.54$\pm$0.17 & 1, 8 \\
{\rm SN}2012{\rm il} & 0.175 & 0.02 & 18.00 (g) & $g\rightarrow 400$ & 0.09 & -0.18 & 18.09$\pm$0.21 & 2.19$\pm$0.16 & 0.49$\pm$0.17 & 1, 8 \\
{\rm PTF}11{\rm rks} & 0.190 & 0.04 & 19.13 (g) & $g\rightarrow 400$ & 0.17 & -0.19 & 19.15$\pm$0.20 & 2.62$\pm$0.14 & 0.95$\pm$0.14 & 1, 8 \\
{\rm SN}2010{\rm gx} & 0.230 & 0.04 & 18.43 (g) & $g\rightarrow 400$ & 0.13 & -0.23 & 18.53$\pm$0.18 & 2.00$\pm$0.19 & 0.34$\pm$0.16 & 2, 8 \\
{\rm SN}2011{\rm kf} & 0.245 & 0.02 & 18.60 (g) & $g\rightarrow 400$ & 0.09 & -0.13 & 18.64$\pm$0.18 & 1.49$\pm$0.16 & - & 1, 8 \\
{\rm LSQ}12{\rm dlf} & 0.255 & 0.01 & 18.78 (V) & $V\rightarrow 400$ & 0.03 & -0.27 & 19.02$\pm$0.12 & 1.00$\pm$0.10 & 0.29$\pm$0.12 & 3, 8 \\
{\rm PTF}09{\rm cnd} & 0.258 & 0.03 & 18.29 (R) & $R\rightarrow 400$ & 0.05 & -0.34 & 18.58$\pm$0.22 & 1.09$\pm$0.14 & - & 4, 8 \\
{\rm SN}2013{\rm dg} & 0.265 & 0.01 & 19.06 (g) & $g\rightarrow 400$ & 0.03 & -0.30 & 19.33$\pm$0.20 & 2.08$\pm$0.20 & 0.51$\pm$0.17 & 3, 8 \\
{\rm PS}1-10{\rm bzj} & 0.650 & 0.01 & 21.23 (r) & $r\rightarrow 400$ & 0.02 & -0.58 & 21.79$\pm$0.20 & 1.70$\pm$0.14 & 0.60$\pm$0.15 & 5, 8 \\
{\rm PS}1-10{\rm ky} & 0.956 & 0.03 & 21.15 (i) & $i\rightarrow 400$ & 0.06 & -0.73 & 21.82$\pm$0.18 & 1.31$\pm$0.15 & 0.30$\pm$0.18 & 6, 8 \\
{\rm SCP}-06{\rm F}6 & 1.189 & 0.01 & 21.04 (z) & $z\rightarrow 400$ & 0.01 & -1.35 & 22.38$\pm$0.20 & 0.89$\pm$0.15 & - & 7, 8 \\
&&&&&&&&&& \\
\enddata
\break\break\break
\vbox{
Notes: All values are from Inserra \& Smartt (2014), except for
$m(400)$, which is calculated here. The error bars are directly from
Figures~5 and 6 in Inserra \& Smartt (2014). Columns: SN~name; measured
redshift; extinction; observed peak magnitude (AB system) and filter;
Galactic extinction in the observed filter; calculation of the synthetic
400nm magnitude from the observed filter; apparent magnitude mapped into
the 400nm passband, and its dispersion; magnitude decrease in 30 restframe
days and its dispersion; the color change between the~400nm and 520nm
synthetic bands at peak and 30 days later, and its dispersion.
References:
(1) Inserra et al. (2013); (2) Pastorello et al. (2010); (3) Nicholl et al. (2014);
(4) Quimby et al. (2011); (5) Lunnan et al. (2013);
(6) Chomiuk et al. (2011); (7) Barbary et al. (2009); (8) Inserra \& Smartt (2014).\hfill\break}
\end{deluxetable}

The SNe listed in Table 1 were located in faint, dwarf galaxies,
and were unlikely to have suffered significant extinction beyond the
reddening induced by interstellar dust in our Galaxy. We here adopt
the reddening corrections from Tables 1 and 2 of Inserra \& Smartt
(2014), who assumed a standard reddening curve with $R_V=A_V/E(B-V)=3.1$.
We also adopt their $K$-corrections and time dilation effects in
order to obtain the absolute rest-frame peak magnitudes. This
step is necessary due to the large ($0.143<z<1.206$) redshift
coverage of the sample. These are listed
in the table, along with the source names, their redshifts, apparent
peak magnitudes (and filters), the magnitude decrease over 30 days,
and the color change between the 400nm and 520nm synthetic (restframe)
bands at peak and 30 days later.

In order to provide consistent, standarized comparative rest frame
properties, the observed apparent magnitudes ($m$ in Table 1) have
been converted into defined, synthetic magnitudes. Since the SLSN Ic
spectrum around 400nm is continuum dominated, Inserra
\& Smartt (2014) defined a synthetic passband with an effective
width of 80nm, centered at wavelength 400nm, having
steep wings and a flat top. In Table 1, this is referred to as the
400nm band. All of the chosen events in this table have sufficient
photometric coverage to allow the $K$-correction to uniformly
map the observed filter's wavelength range to this 400nm bandpass
in the rest frame. The absolute magnitudes are then formally
defined by the relation
\begin{eqnarray}
M_{\rm peak}(400)&=&m(400)-\mu \nonumber \\
\null&=&m_{f}-K^{\rm peak}_{f\rightarrow 400}-A_{f}-\mu \;,
\end{eqnarray}
where $m(400)$ is the apparent magnitude mapped into the
restframe 400nm band, $\mu$ is the distance modulus calculated from
the luminosity distance, $m_f$ is the AB magnitude in the observed filter $f$
($g$, $V$, $R$, $r$, $z$, or $i$, as indicated in Table~1),
$A_{f}$ is the Galactic extinction in the observed filter,
and $K^{\rm peak}_{f\rightarrow 400}$
is the $K$-correction from the observed filter in Table~1 to the synthetic
400nm bandpass. The values of $A_{f}$ and $m(400)$ are listed
in Table~1. Note that $m(400)$ is a cosmology-independent
apparent magnitude.

The peak magnitude-decline relation for SLSN Ic in the rest frame
400nm band is (Inserra \& Smartt 2014)
\begin{equation}
M_{\rm peak}(400)=M_0+\alpha\Delta M_{30}(400)\;,
\end{equation}
where $\alpha$ is the slope, $\Delta M_{30}(400)$ is the decline
at 30 days, and $M_{0}$ is a constant representing the absolute peak magnitude
at $\Delta M_{30}(400)=0$. Inserra \& Smartt (2014) also found that
$M_{\rm peak}(400)$ appears to have a strong color dependence. Redder objects
are fainter and also become redder faster. This peak magnitude-color evolution
relation is given by
\begin{equation}
M_{\rm peak}(400)=M_0+\alpha\Delta M_{30}(400-520)\;,
\end{equation}
where $\Delta M_{30}(400-520)$ is the difference in color $M(400)-M(520)$
between the peak and 30 days later. (Note that $M_0$ and $\alpha$ need not be the
same in these two expressions.)

With Equations~(2) and (3), an effective (standardized) apparent magnitude $m^{\rm eff}$ may
be obtained as follows:
\begin{equation}
m^{\rm eff}\equiv m(400)-\alpha\Delta M_{30}\;.
\end{equation}
The term $\alpha\Delta M_{30}$ represents an adjustment due to the peak magnitude-decline
relation, in terms of $\alpha\Delta M_{30}(400)$, or the peak magnitude-color evolution
relation, in terms of $\alpha\Delta M_{30}(400-520)$, as the case may be.
The effective apparent magnitude may also be expressed as (Perlmutter et al. 1997,1999)
\begin{equation}
m^{\rm eff}=\Upsilon + 5\log_{10}[H_0\,d_L(z)] \;,
\end{equation}
where $H_0$ is the Hubble constant in units of km $\rm s^{-1}$ $\rm Mpc^{-1}$
and $d_L$ is the luminosity distance in units of Mpc.
Here $\Upsilon$ is the ``$H_0$-free" 400nm absolute
peak magnitude, represented in terms of the standardizable absolute
magnitude $M_0$, according to the definition
\begin{equation}
\Upsilon\equiv M_0-5\log_{10}(H_0)+25
\end{equation}
(see Li et al. 2014 and references cited therein).

The apparent correlations suggest the use of two ``nuisance" parameters
($\alpha$ and $\Upsilon$) whose optimization along with the model
parameters decreases the overall scatter in the distance
modulus. For each model, we therefore find the best fit by
minimizing the $\chi^2$ statistic, defined as follows:

\begin{equation}
\chi^2=\sum_{i} {{\left[m_{i}(400)
-\alpha\Delta M_{30,\,i} -\Upsilon -5\log_{10}[H_0\,d_L(z_i)]\right]}^2 \over
\sigma_{m_{i}(400)}^2+\alpha^2\sigma_{\Delta M_{30,\,i}}^2}\;.
\end{equation}

 In $\Lambda$CDM, the luminosity distance $d_L^{\Lambda{\rm CDM}}$ depends
on several parameters, including $H_0$ and the mass fractions
$\Omega_{\rm m} \equiv \rho_{\rm m}/\rho_{\rm c}$, $\Omega_{\rm r}\equiv
\rho_{\rm r}/\rho_{\rm c}$, and $\Omega_{\rm de}\equiv \rho_{\rm de}/
\rho_{\rm c}$, defined in terms of the current matter ($\rho_{\rm m}$),
radiation ($\rho_{\rm r}$), and dark-energy ($\rho_{\rm de}$) densities,
and the critical density $\rho_{\rm c}\equiv 3c^2H_0^2/8\pi G$.
Assuming zero spatial curvature, so that $\Omega_{\rm m}+\Omega_{\rm r}
+\Omega_{\rm de}=1$, the luminosity distance to redshift $z$ is given by the expression
\begin{equation}
d_L^{\Lambda{\rm CDM}}(z)={c\over H_0}(1+z)\int_0^z
\left[\Omega_{\rm m}(1+z)^3+\Omega_{\rm r}(1+z)^4+\Omega_{\rm de}
(1+z)^{3(1+w_{\rm de})}\right]^{-1/2}\;dz\;,
\end{equation}
where $p_{\rm de}=w_{\rm de}\rho_{\rm de}$ is the dark-energy equation
of state. The Hubble constant $H_0$ cancels out in Equation~(7) when we
multiply $d_L$ by $H_0$, so the essential remaining parameters in flat
$\Lambda$CDM are $\Omega_{\rm m}$ and $w_{\rm de}$. If we further
assume that dark energy is a cosmological constant with $w_{\rm de}=-1$,
then only the parameter $\Omega_{\rm m}$ is available to fit the data.

In the $R_{\rm h}=ct$ Universe (Melia 2007; Melia \& Abdelqader 2009; Melia \& Shevchuk 2012),
the luminosity distance depends only on $H_0$, but since here too the Hubble constant cancels
out in the product $H_0d_L$, there are actually no free (model) parameters left to fit the SLSN Ic
data. In this cosmology,
\begin{equation}
d_L^{R_{\rm h}=ct}(z)={c\over H_0}(1+z)\ln(1+z)\;.
\end{equation}

\section{Results}
We have assumed that SLSNe Ic can be used as standardizable candles and applied the
$\Delta M_{30}$ decline relation (with 11 objects) and the peak magnitude-color
evolution relation (with 8 objects) to compare the standard ($\Lambda$CDM) model
with the $R_{\rm h}=ct$ Universe. In this section, we discuss how the fits have been
optimized, first for $\Lambda$CDM, and then for $R_{\rm h}=ct$. The outcome for each
model is more fully described and discussed in subsequent sections.

\subsection{$\Lambda$CDM}
In the most basic $\Lambda$CDM model, the dark-energy equation of state parameter,
$w_{\rm de}$, is exactly $-1$. The Hubble constant $H_0$ cancels out in Equation~(7) when we
multiply $d_L$ by $H_0$, so the essential remaining parameter
is $\Omega_{\rm m}$. Type Ia SN measurements (see, e.g.,
Perlmutter et al.  1998, 1999; Riess et al. 1998; Schmidt et al. 1998),
CMB anisotropy data (e.g., Hinshaw et al. 2013), and baryon acoustic oscillation (BAO)
peak length scale estimates (e.g., Samushia \& Ratra 2009),
strongly suggest that we live in a spatially flat, dark energy-dominated universe
with concordance parameter values $\Omega_{\rm m}\approx0.27$ and
$H_{0}\approx70$ km $\rm s^{-1}$ $\rm Mpc^{-1}$.

\begin{figure}[h]
\centerline{\includegraphics[angle=0,scale=1.0]{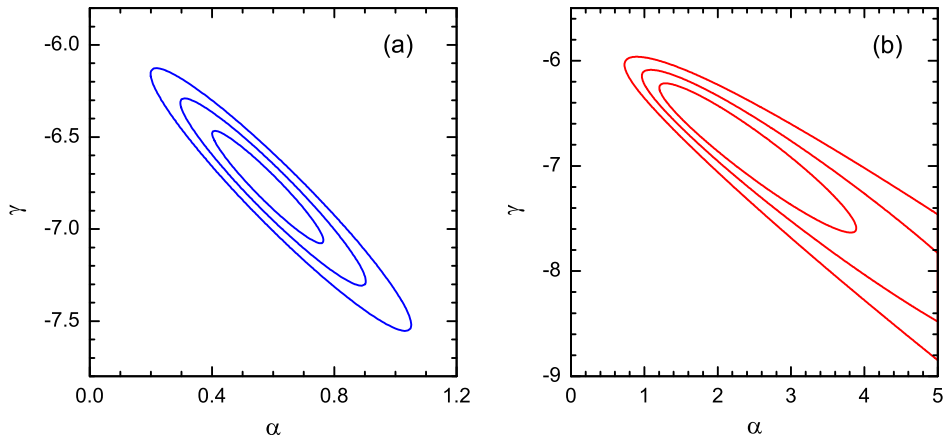}}
\caption{$1$-$3\sigma$ constraints on $\alpha$ and $\Upsilon$ for the concordance model,
using the $\Delta M_{30}$ decline relation (panel a) and the peak magnitude-color evolution relation
(panel b).}\label{concordance}
\end{figure}

We will therefore first attempt to fit the data with this concordance model, using prior values
for all the model parameters, but not the two ``nuisance" parameters $\alpha$ and $\Upsilon$.
For the $\Delta M_{30}$ decline relation (with 11 objects), the resulting constraints
on $\alpha$ and $\Upsilon$ are shown in Figure~1(a). For this fit, we obtain
$\alpha=0.57^{+0.19}_{-0.17}$ $(1\sigma)$, $\Upsilon=-6.75^{+0.28}_{-0.32}$ $(1\sigma)$,
and a $\chi^{2}$ per degree of freedom of $\chi^2_{\rm dof}=27.04/9=3.00$, remembering that all
of the $\Lambda$CDM parameters are assumed to have prior values, except for the
``nuisance" parameters $\alpha$ and $\Upsilon$. (The corresponding data and best
fit are shown in the top lefthand panel of Figure~4.) For the peak magnitude-color
evolution relation (with 8 objects), the resulting constraints on $\alpha$ and
$\Upsilon$ are shown in Figure~1(b). The best-fit parameters are
$\alpha=2.00^{+1.91}_{-0.79}$ $(1\sigma)$, $\Upsilon=-6.65^{+0.45}_{-0.97}$ $(1\sigma)$.
The $\chi^{2}$ per degree of freedom for the concordance model with these optimized
``nuisance" parameters is $\chi^2_{\rm dof}=2.48/6=0.41$ (see also the top righthand
panel in Figure~4).

\begin{figure}[hp]
\vskip-0.3in
\centerline{\includegraphics[angle=0,scale=0.8]{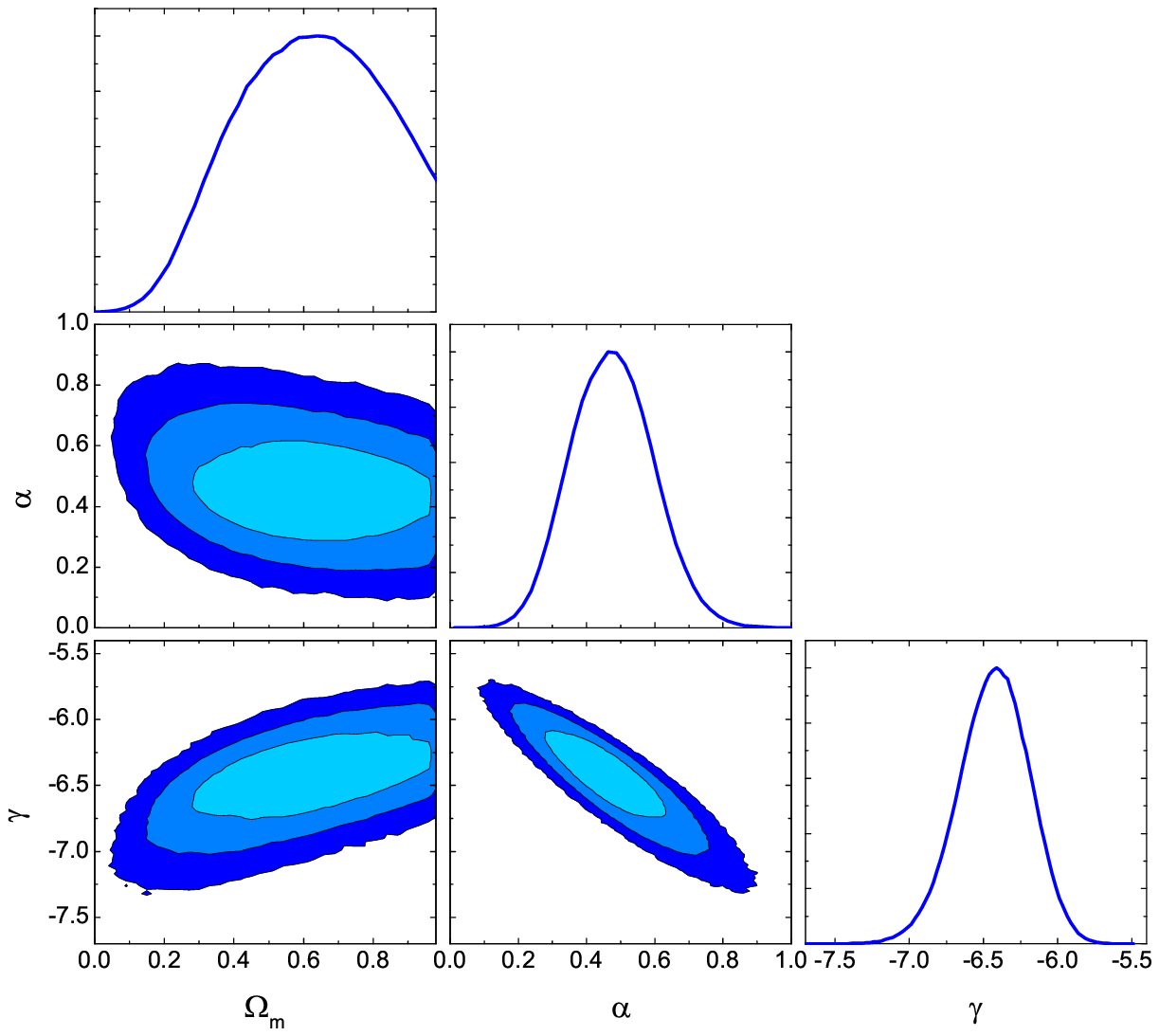}}
\vskip -0.4in
\centerline{\includegraphics[angle=0,scale=0.8]{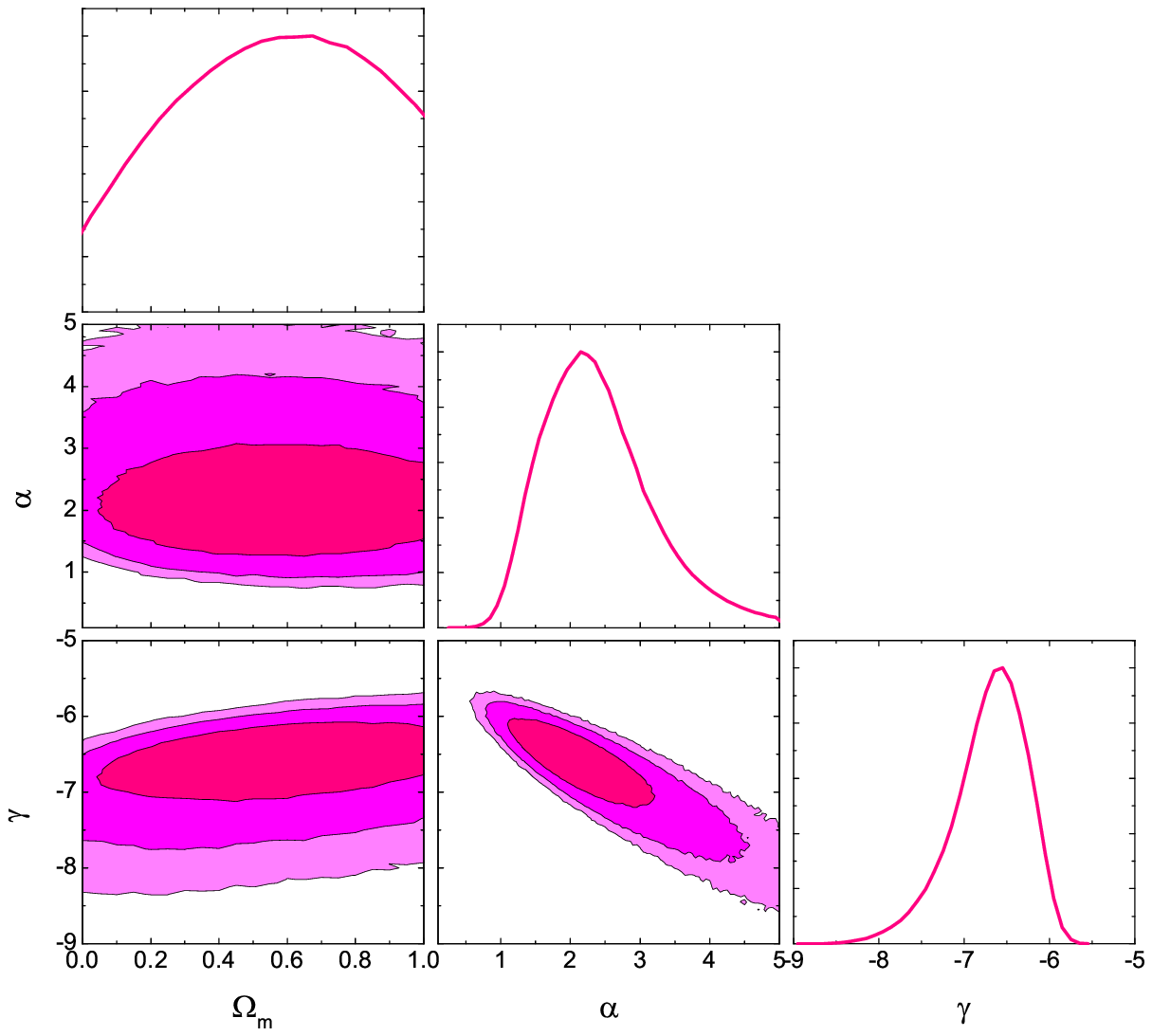}}
\vskip-0.2in
\caption{Top panel: Constraints on $\alpha$, $\Upsilon$ and
$\Omega_{\rm m}$ for the flat $\Lambda$CDM model, using
the $\Delta M_{30}$ decline relation. Bottom panel: Same as the top panel,
but using the peak magnitude-color evolution relation.}\label{LCDM}
\end{figure}

If we relax some of the priors, and allow $\Omega_{\rm m}$ to be a free parameter,
we obtain the $1$-$3\sigma$ constraint contours shown in the top panel of Figure~2
for the $\Delta M_{30}$ decline relation, and the bottom panel for the peak
magnitude-color evolution relation. These contours show that at the $1\sigma$ level,
the optimized parameter values are $\alpha=0.46^{+0.18}_{-0.18}$, $\Upsilon=
-6.41^{+0.34}_{-0.33}$, and $\Omega_{m}=0.62^{+0.35}_{-0.34}$. We find
that the $\chi^{2}$ per degree of freedom for the optimized flat $\Lambda$CDM
model is $\chi^2_{\rm dof}=24.89/8=3.11$. The bottom panel of Figure~2 shows
the $1$-$3\sigma$ constraints on $\alpha$, $\Upsilon$, and $\Omega_{\rm m}$
for the flat $\Lambda$CDM model, using the peak magnitude-color evolution relation.
The contours show that at the $1\sigma$-level, $\alpha=2.05^{+1.19}_{-0.94}$,
$\Upsilon=-6.52^{+0.50}_{-0.68}$, but that $\Omega_{\rm m}$ is poorly
constrained; only a lower limit of $\sim0.05$ can be set at this confidence
level and the best-fit $\Omega_{\rm m}$ is 0.59. The $\chi^{2}$ per
degree of freedom is $\chi^2_{\rm dof}=2.15/5=0.43$.

\subsection{The $R_{\rm h}=ct$ Universe}
The $R_{\rm h}=ct$ Universe has only one free parameter, $H_{0}$,
but since the Hubble constant cancels out in the product $H_{0}d_L$,
there are actually no free (model) parameters left to fit the SLSN Ic data.
The results of fitting the $\Delta M_{30}$ decline relation with this cosmology
are shown in Figure~3(a). We see here that the best fit corresponds to
$\alpha=0.50^{+0.19}_{-0.17}$ $(1\sigma)$ and $\Upsilon=-6.48^{+0.28}_{-0.31}$ $(1\sigma)$.
With $11-2=9$ degrees of freedom, the reduced $\chi^{2}$ is $\chi_{\rm dof}^{2}=
25.79/9=2.87$. The results of fitting the peak magnitude-color evolution relation
with this cosmology are shown in Figure~3(b). We see here that the best fit corresponds to
$\alpha=1.88^{+1.84}_{-0.78}$ $(1\sigma)$ and $\Upsilon=-6.44^{+0.44}_{-0.94}$ $(1\sigma)$.
With $8-2=6$ degrees of freedom, the reduced $\chi^{2}$ is $\chi_{\rm dof}^{2}=
2.19/6=0.37$.

\begin{figure}[h]
\centerline{\includegraphics[angle=0,scale=1.0]{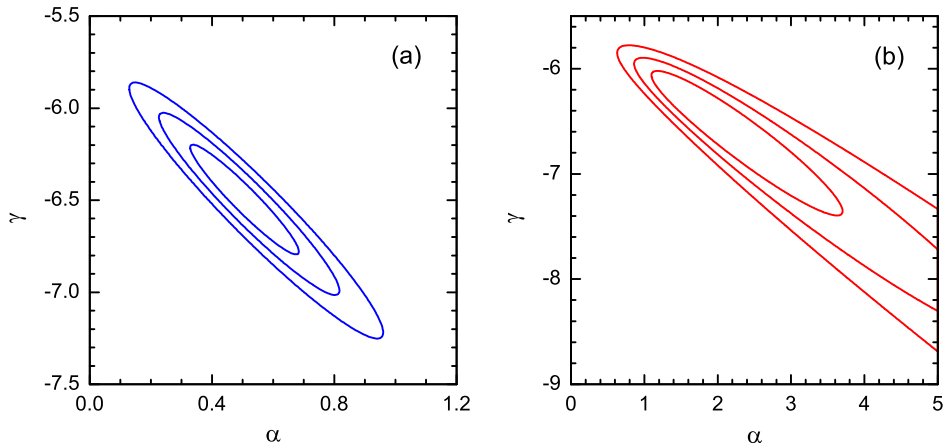}}
\caption{$1$-$3\sigma$ constraints on $\alpha$ and $\Upsilon$ for the $R_{\rm h}=ct$ Universe,
using the $\Delta M_{30}$ decline relation (panel a) and the peak magnitude-color evolution relation
(panel b).}\label{Rh}
\end{figure}

\begin{figure*}
\hskip -0.04in
\includegraphics[angle=0,scale=0.55]{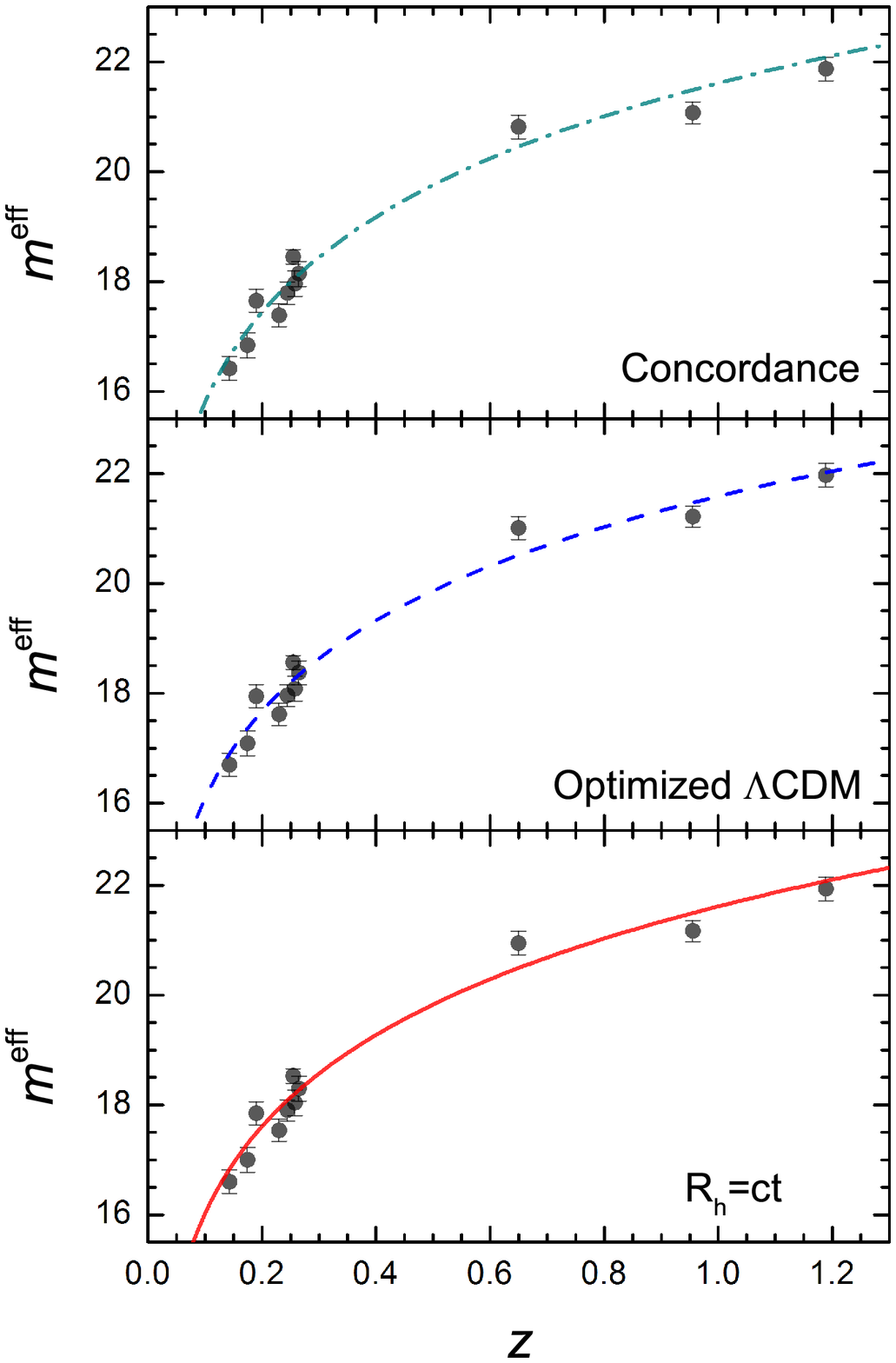}\hskip 0.2in
\includegraphics[angle=0,scale=0.55]{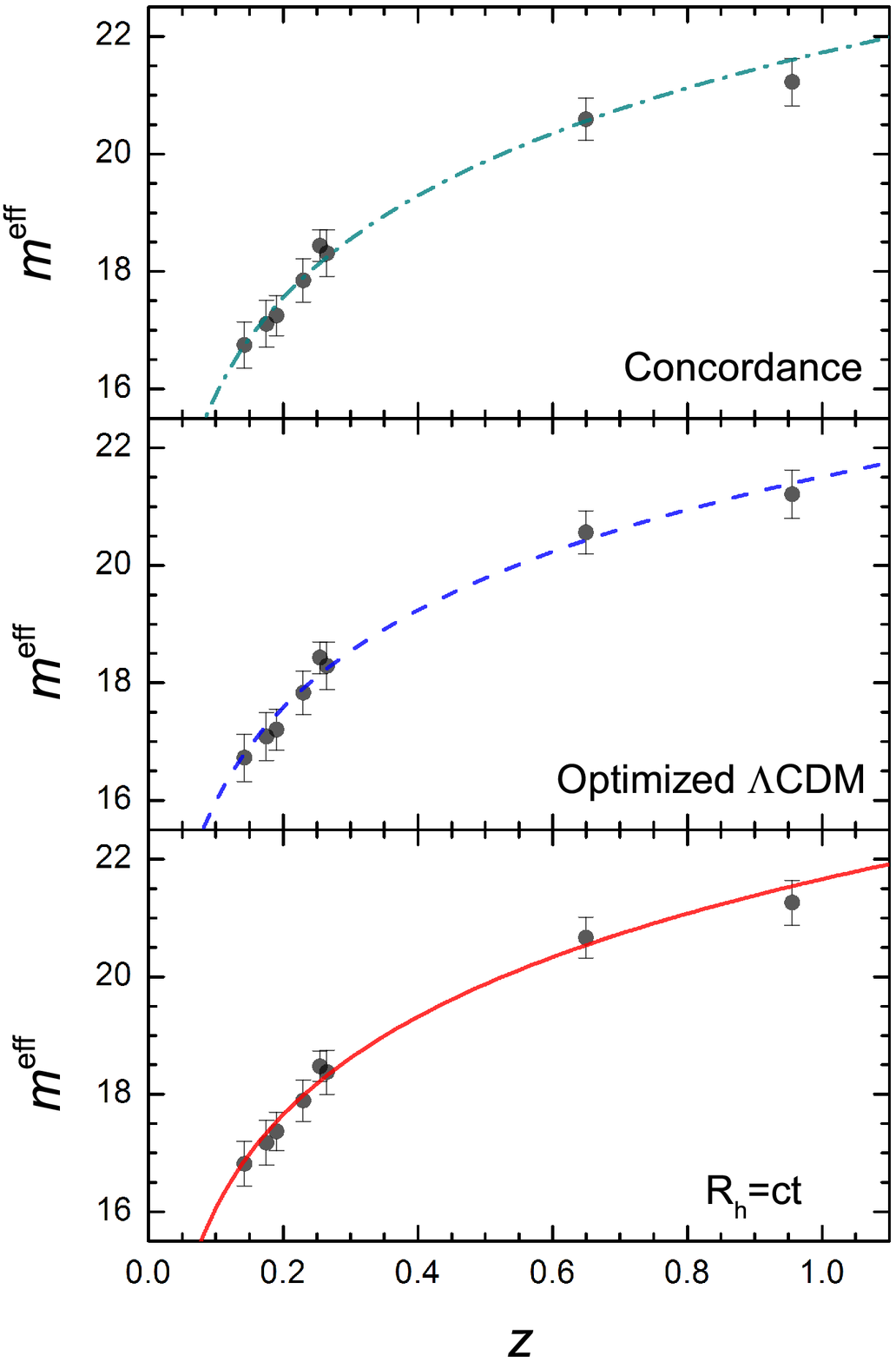}
    \caption{Hubble diagrams for SLSN Ic. The effective apparent magnitudes of SLSN Ic
    are plotted as solid points, together with their corresponding best-fit theoretical
    curves. Left: The effective magnitudes are calculated using the $\Delta M_{30}$ decline
    relation. Right: The effective magnitudes are calculated using the peak magnitude-color
    evolution relation.}\label{HD}
\end{figure*}

To facilitate a direct comparison between $\Lambda$CDM and $R_{\rm h}=ct$,
we show in Figure~4 the Hubble diagrams for SLSNe Ic. In the left panel of
Figure~4, the effective magnitudes $m^{\rm eff}$ of 11 SLSN Ic are plotted
as solid points, together with the best-fit theoretical curves (from top to bottom)
for the concordance model (with prior values of the parameters, and with
$\alpha=0.57$ and $\Upsilon=-6.75$), for the  optimized flat $\Lambda$CDM model
(with $\Omega_{\rm m}=0.62$, and with $\alpha=0.46$, and $\Upsilon=-6.41$).
and for the $R_{\rm h}=ct$ Universe (with $\alpha=0.50$ and $\Upsilon=-6.48$).
The Hubble diagrams derived using the peak magnitude-color evolution
relation are shown in the righthand panels of Figure~4. Here, the effective
magnitudes $m^{\rm eff}$ of 8 SLSN Ic are plotted as solid points, together
with the best-fit theoretical curves for the concordance model (with prior
values of the parameters, and with $\alpha=2.00$ and $\Upsilon=-6.65$),
for the optimized flat $\Lambda$CDM model (with $\Omega_{\rm m}=0.59$,
and with $\alpha=2.05$, and $\Upsilon=-6.52$), and for the $R_{\rm h}=ct$
Universe (with $\alpha=1.88$ and $\Upsilon=-6.44$). Strictly based on their
$\chi^{2}_{\rm dof}$ values, the optimized $\Lambda$CDM model
and the $R_{\rm  h}=ct$ Universe appear to fit the SLSN Ic data comparably well.
However, because these models formulate their observables (such as the
luminosity distances in Equations~8 and 9) differently, and because they
do not have the same number of free parameters, a comparison of the likelihoods
for either being closer to the `true' model must be based on model selection tools.

\subsection{Model Selection Tools}
Several information criteria commonly used in cosmology (see,
e.g., Melia \& Maier 2013, and references cited therein) include
the Akaike Information Criterion, ${\rm AIC}=\chi^{2}+2n$, where
$n$ is the number of free parameters (Akaike 1973; Takeuchi 2000;
Liddle 2004, 2007; Tan \& Biswas 2012), the Kullback Information
Criterion, ${\rm KIC}=\chi^{2}+3n$ (Bhansali \& Downnham 1977;
Cavanaugh 1999, 2004), and the Bayes Information Criterion,
${\rm BIC}=\chi^{2}+(\ln N)n$, where $N$ is the number of data points
(Schwarz 1978; Liddle et al. 2006; Liddle 2007). With ${\rm AIC}_\alpha$
characterizing model $\mathcal{M}_\alpha$,
the unnormalized confidence that this model is true is the Akaike
weight $\exp(-{\rm AIC}_\alpha/2)$. Model $\mathcal{M}_\alpha$ has likelihood
\begin{equation}
P(\mathcal{M}_\alpha)= \frac{\exp(-{\rm AIC}_\alpha/2)}
{\exp(-{\rm AIC}_1/2)+\exp(-{\rm AIC}_2/2)}
\end{equation}
of being the correct choice in this one-on-one comparison. The difference
$\Delta \rm AIC \equiv {\rm AIC}_2\nobreak-{\rm AIC}_1$ determines the
extent to which $\mathcal{M}_1$ is favoured over~$\mathcal{M}_2$. For
Kullback and Bayes, the likelihoods are defined analogously.

With the optimized fits we have reported in this paper, our analysis of the
$\Delta M_{30}$ decline relation (with 11 objects) shows that
$R_{\rm h}=ct$ is favoured over the flat $\Lambda$CDM model with a likelihood of
$\approx 63.4\%$ versus $36.6\%$ using AIC, $\approx 74.1\%$ versus $\approx 25.9\%$
using KIC, and $\approx 67.9\%$ versus $\approx 32.1\%$ using BIC.
In our one-on-one comparison using the peak magnitude-color
evolution relation (with 8 objects), the $R_{\rm h}=ct$ Universe
is preferred over $\Lambda$CDM with a likelihood of $\approx 72.7\%$ versus
$27.3\%$ using AIC, $\approx 81.5\%$ versus $\approx 18.5\%$
using KIC, and $\approx 73.5\%$ versus $\approx 26.5\%$ using BIC.

\section{Monte Carlo Simulations with a Mock Sample}
These results are interesting, though the current sample of SLSNe Ic is clearly
too small for either model to be ruled out just yet. However, this situation will
change with the discovery of new SNe, particularly at redshifts $z> 2$.
To anticipate how well the SLSN Ic catalog obeying the peak magnitude-color
evolution relation may be used to constrain the dark-energy equation of state in
$\Lambda$CDM, and to differentiate between this standard model and the
$R_{\rm h}=ct$ Universe, we will here produce mock samples of SLSNe Ic
based on the current measurement accuracy. In using the model selection tools,
the outcome $\Delta\equiv$ AIC$_1-$ AIC$_2$ (and analogously for KIC and BIC)
is judged `positive' in the range $\Delta=2-6$, `strong' for $\Delta=6-10$,
and `very strong' for $\Delta>10$. In this section, we will estimate the sample
size required to significantly strengthen the evidence in favour of $R_{\rm h}=ct$
or $\Lambda$CDM, by conservatively seeking an outcome even beyond $\Delta\simeq11.62$,
i.e., we will see what is required to produce a likelihood $\sim 99.7\%$ versus
$\sim 0.3\%$, corresponding to a $3\sigma$ confidence level.

We will consider two cases: one in which the background cosmology is
assumed to be $\Lambda$CDM, and a second in which it is $R_{\rm h}=ct$,
and we will attempt to estimate the number of SLSNe Ic required in each case
to rule out the alternative (presumably incorrect) model at a $\sim 99.7\%$
confidence level. The synthetic SLSNe Ic are each characterized by a set of
parameters denoted as ($z$, $m[400]$, $\Delta M_{30}[400-520]$).
We generate the synthetic sample using the following procedure:

1. Since a subclass of broad-lined Type Ic SNe are observed to be associated with
Gamma-ray bursts (GRBs; Hjorth et al. 2003; Stanek et al. 2003), we simulate the
\emph{z}-distribution of our sample based on the observed \emph{z}-distribution
of GRBs. Shao et al. (2011) found that the \emph{z}-distribution of GRBs appears to be
asymptotic to a parameter-free probability density distribution, $f(z)=ze^{-z}$.
The redshift $z$ of our SLSN Ic events is generated randomly from this function.
Since they can be discovered out to redshifts $z\sim4$ (e.g., Chomiuk et al. 2011;
Berger et al. 2012; Cooke et al. 2012; Howell et al. 2013),
the range of redshifts for our analysis is $[0 ,4]$. We assign the absolute
peak magnitude $M(400)$ uniformly between -23.0 and -21.0 mag, based on
the current SLSN Ic measurements.

2. The synthetic apparent peak magnitude $m(400)$ is calculated using
the relation $m(400)=M(400)+\mu(z)$, where $\mu(z)$ is the distance
modulus at $z$, for a flat $\Lambda$CDM cosmology with $\Omega_{\rm m}=
0.27$ and $H_{0}=70$ km $\rm s^{-1}$ $\rm Mpc^{-1}$ (\S~4.1), or the $R_{\rm h}=ct$
Universe with $H_{0}=70$ km $\rm s^{-1}$ $\rm Mpc^{-1}$ (\S~4.2).

3. With the selected $z$ and $M(400)$ values, we first infer a
$\Delta M_{30}(400-520)$ from the peak magnitude-color evolution
relation in a flat $\Lambda$CDM cosmology with $\Omega_{\rm m}=0.27$
and $H_{0}=70$ km $\rm s^{-1}$ $\rm Mpc^{-1}$ (\S~4.1), or the $R_{\rm h}=ct$
Universe with $H_{0}=70$ km $\rm s^{-1}$ $\rm Mpc^{-1}$ (\S~4.2). And then
we add scatter to this relation by assigning a dispersion to the
$\Delta M_{30}(400-520)$ value;  i.e., we randomly map this quantity
to the new value $\Delta M^\prime_{30}(400-520)$ assuming a normal distribution
with a center at $\Delta M_{30}(400-520)$ and a dispersion $\sigma=0.1$ mag.
This value of $\sigma$ is typical for the current (observed) peak magnitude-color
evolution relation, which yields a standard deviation $\sigma=0.08$ mag
for the linear fit (Inserra \& Smartt 2014). If $\Delta M^\prime_{30}(400-520)>0.0$ mag,
this SLSN Ic is included in our sample. Otherwise, it is excluded.

4. We next assign ``observational" errors to $m(400)$ and $\Delta M^\prime_{30}(400-520)$.
We will assign a dispersion $\sigma_{i}=0.05+0.05k_i$ to each event $i$,
where $k_i$ is a random number between 0 and 1.

This sequence of steps is repeated for each new SLSN Ic in the sample, until
we reach the likelihood criterion discussed above. As with the real 11-SLSN Ic
sample, we optimize the model fits by minimizing the $\chi^{2}$ function
in Equation~(7).

\subsection{Assuming $\Lambda$CDM as the Background Cosmology}
We have found that a sample of at least 240 SLSNe Ic is required
in order to rule out $R_{\rm h}=ct$ at the $\sim 99.7 \%$ confidence level. The
optimized parameters corresponding to the best-fit $w$CDM model for
these simulated data are displayed in Figure~5. To allow for the greatest
flexibility in this fit, we relax the assumption that dark energy is a cosmological
constant with $w_{\rm de}=-1$, and allow $w_{\rm de}$ to be a free parameter,
along with $\Omega_{\rm m}$. Figure~5 shows the 1-D probability distribution
for each parameter ($\Omega_{\rm m}$, $w_{\rm de}$, $\alpha$, $\Upsilon$),
and 2-D plots of the $1$-$3\sigma$ confidence regions for two-parameter combinations.
The best-fit values for $w$CDM using the simulated sample with 240 SNe
in the $\Lambda$CDM model are $\Omega_{\rm m}=0.28_{-0.04}^{+0.03}$ $(1\sigma)$,
$w_{\rm de}=-1.14_{-0.35}^{+0.28}$ $(1\sigma)$, $\alpha=2.21_{-0.10}^{+0.10}$ $(1\sigma)$,
and $\Upsilon=-6.80_{-0.13}^{+0.10}$ $(1\sigma)$.

\begin{figure}[hp]
\vskip-0.3in
\centerline{\includegraphics[angle=0,scale=1.2]{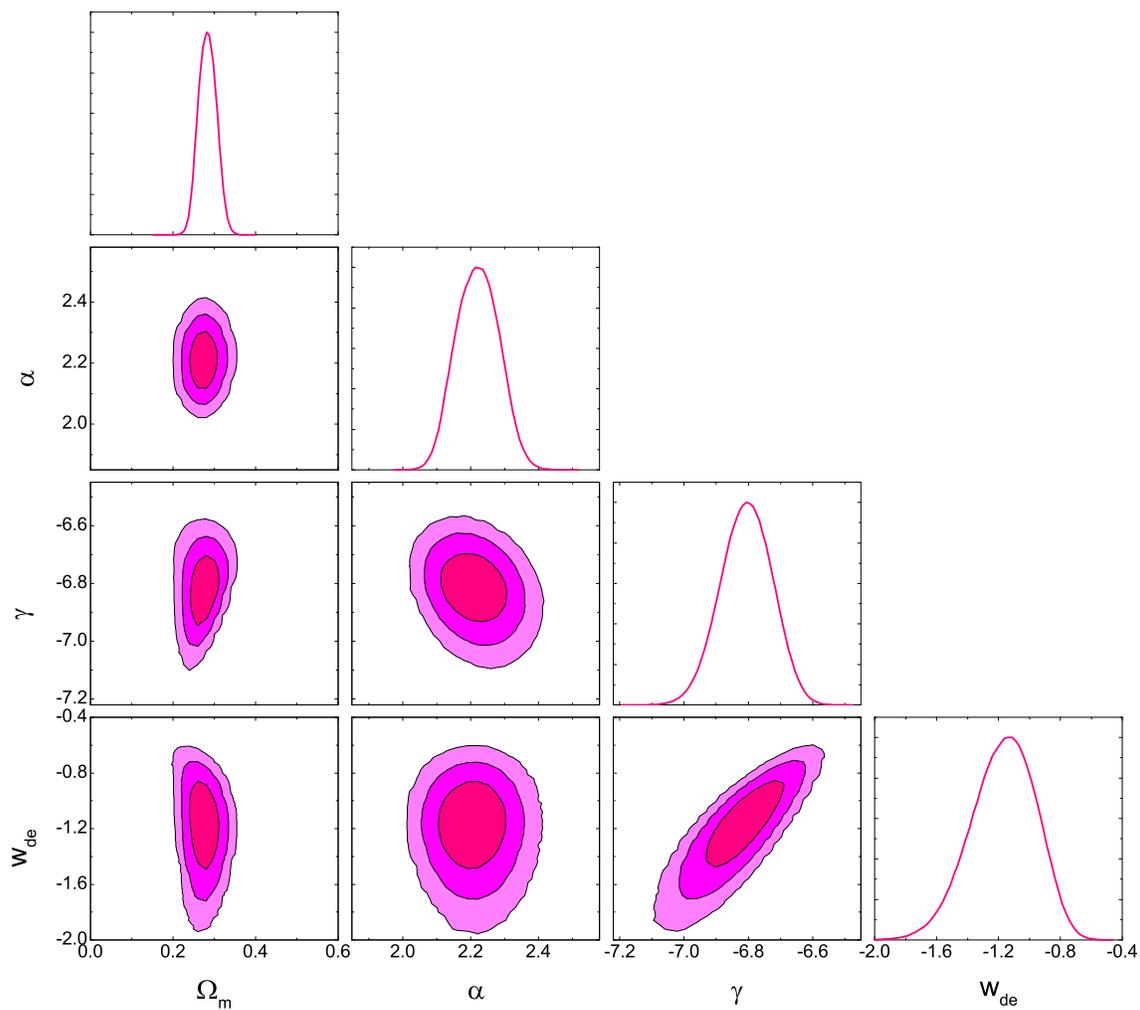}}
\vskip-0.1in
\caption{The 1-D probability distributions and 2-D regions with the $1$-$3\sigma$
contours corresponding to the parameters $\Omega_{\rm m}$, $w_{\rm de}$, $\alpha$,
and $\Upsilon$ in the best-fit $w$CDM model, using the simulated sample with
240 SLSNe Ic, assuming $\Lambda$CDM as the background cosmology.}
\end{figure}

To gauge the impact of these constraints more clearly, we show in Figure~6 the confidence
regions (shaded, with red contours) for $\Omega_{\rm m}$ and $w_{\rm de}$ using these
240 simulated SLSNe Ic (the same as the bottom left-hand panel of Figure~5), and compare
these to the constraint contours for the 580 Union2.1 Type Ia SN data (Suzuki et al. 2012)
(represented by the blue contours in Figure~6). It is straightforward to see how effectively
the SLSNe Ic could be used as a cosmological tool, because the confidence regions resulting
from their analysis are smaller (and narrower for $\Omega_{\rm m}$) than those
corresponding to the Type Ia events. The better constraints are mainly due to the fact that
SLSNe Ic are distributed over a much wider redshift range, extending towards high-\emph{z},
where tighter constraints on the model can be achieved.

\begin{figure}[h]
\vskip 0.1in
\centerline{\includegraphics[angle=0,scale=1.0]{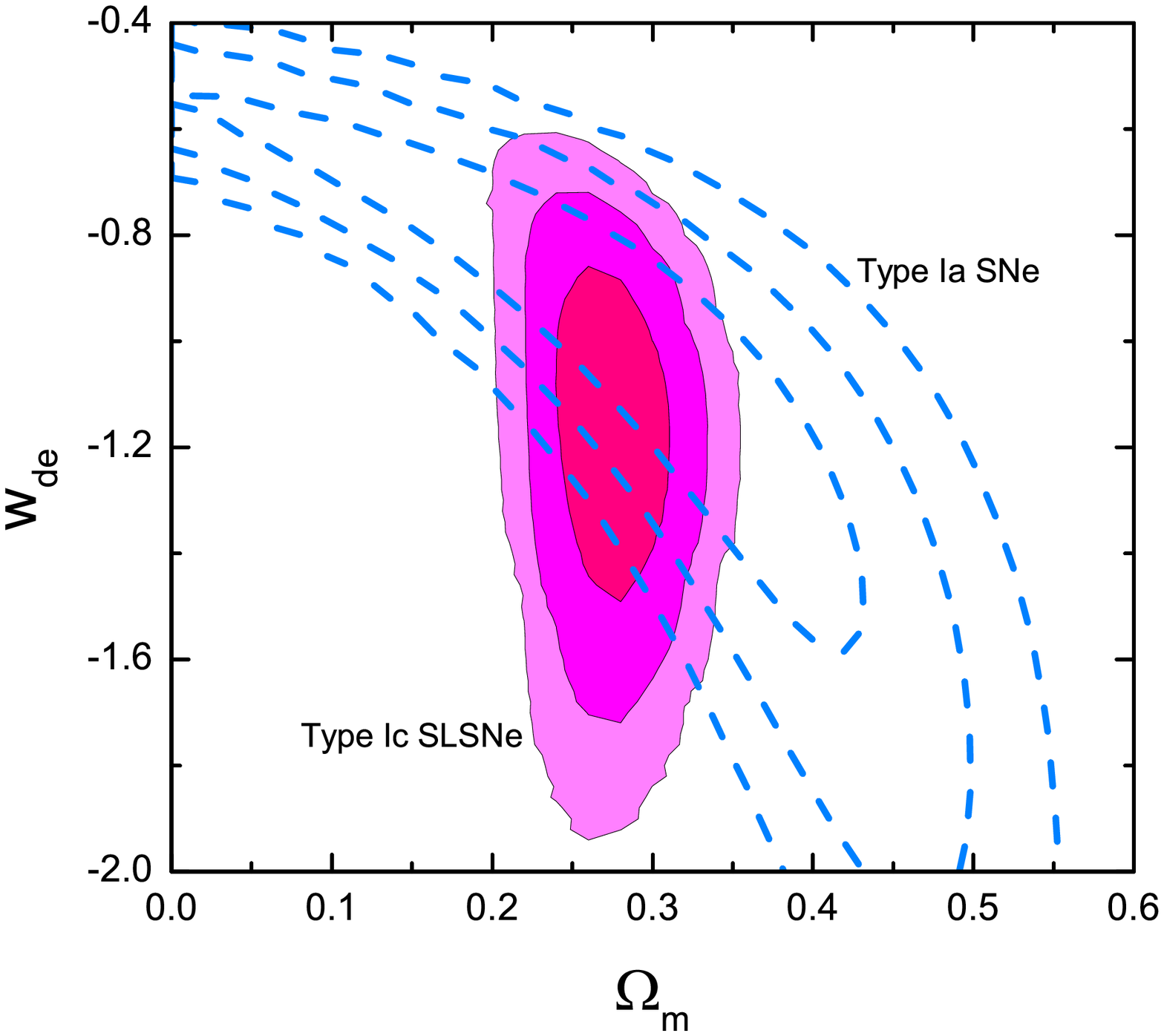}}
\caption{One-$\sigma$, 2-$\sigma$ and 3-$\sigma$ confidence regions
for the $w$CDM model using simulated SLSNe Ic (shaded, with red contours),
compared with those (dashed blue contours) associated with the 580 Union2.1
Type Ia SNe.}
\end{figure}

In Figure~7, we show the corresponding 2-D contours in the $\alpha-\Upsilon$ plane for
the $R_{\rm h}=ct$ Universe. The best-fit values for the simulated sample are
$\alpha=2.22_{-0.08}^{+0.09}$ $(1\sigma)$ and $\Upsilon=-6.53_{-0.04}^{+0.03}$ $(1\sigma)$.

\begin{figure}[h]
\vskip -0.2in
\centerline{\hskip 1.1in\includegraphics[angle=0,scale=1.0]{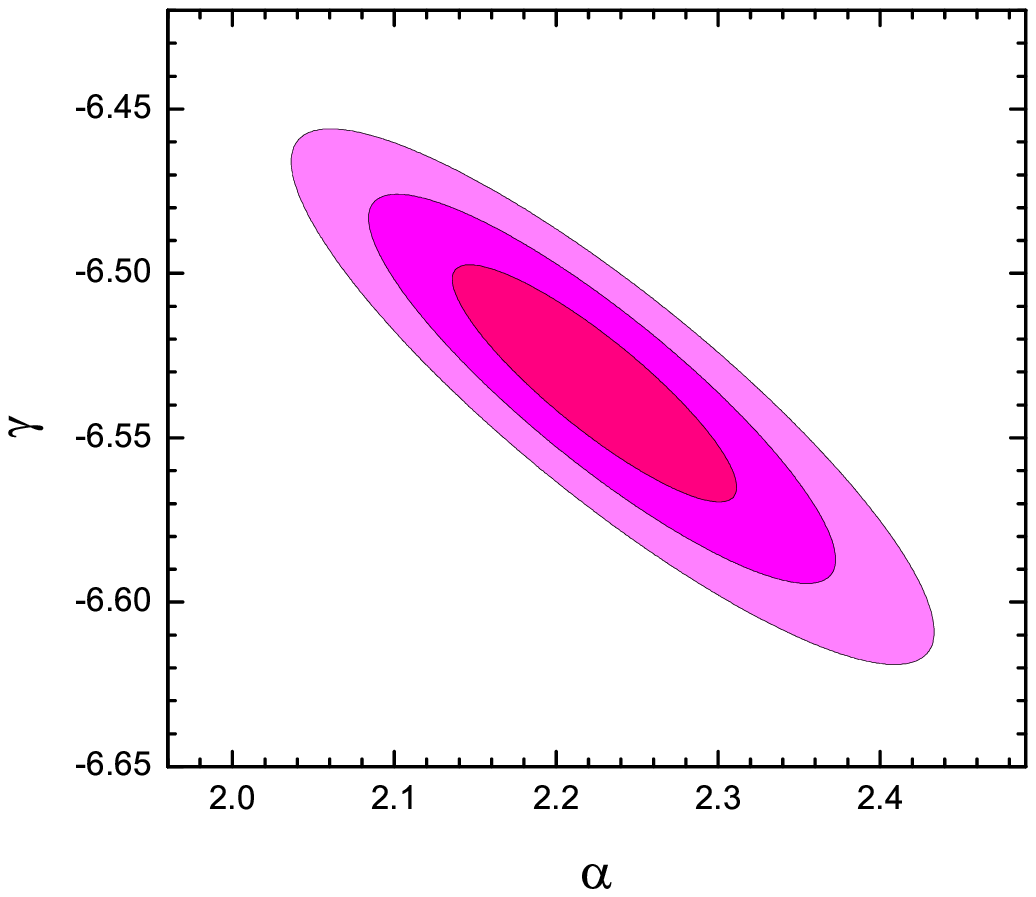}}
\caption{The 2-D region with the $1$-$3\sigma$
contours for the parameters $\alpha$ and $\Upsilon$ in the $R_{\rm h}=ct$ Universe,
using a sample of 240 SLSN Ic, simulated with $\Lambda$CDM as the
background cosmology. The simulated model parameters were $\Omega_{\rm m}=0.27$,
$\Omega_{\Lambda}=0.73$, and $H_{0}=70$ km $\rm s^{-1}$ $\rm Mpc^{-1}$.}
\end{figure}

Since the number $N$ of data points in the sample is now much greater than one, the
most appropriate information criterion to use is the BIC. The logarithmic penalty
in this model selection tool strongly suppresses overfitting if $N$ is large
(the situation we have here, which is deep in the asymptotic regime). With $N=240$,
our analysis of the simulated sample shows that the BIC would favor the $w$CDM
model over $R_{\rm h}=ct$ by an overwhelming likelihood of $99.7\%$ versus only $0.3\%$
(i.e., the prescribed $3\sigma$ confidence limit).

\subsection{Assuming $R_{\rm h}=ct$ as the Background Cosmology}
In this case, we assume that the background cosmology is the $R_{\rm h}=ct$ Universe,
and seek the minimum sample size to rule out $w$CDM at the
$3\sigma$ confidence level. We have found that a minimum of 480 SLSNe Ic
are required to achieve this goal. To allow for the greatest flexibility
in the $w$CDM fit, here too we relax the assumption of dark energy as a cosmological
constant with $w_{\rm de}=-1$, and allow $w_{\rm de}$ to be a free parameter,
along with $\Omega_{\rm m}$. In Figure~8, we show the 1-D probability distribution
for each parameter ($\Omega_{\rm m}$, $w_{\rm de}$, $\alpha$, $\Upsilon$), and 2-D plots of
the $1$-$3\sigma$ confidence regions for two-parameter combinations. The best-fit values
for $w$CDM using this simulated sample with 480 SLSNe Ic are $\Omega_{\rm m}=0.0$,
$w_{\rm de}=-0.33_{-0.06}^{+0.01}$ $(1\sigma)$, $\alpha=2.05_{-0.07}^{+0.07}$
$(1\sigma)$, and $\Upsilon=-6.52_{-0.05}^{+0.04}$ $(1\sigma)$.
Note that the simulated SLSNe Ic give a good constraint on $w_{\rm de}$, but
a weak constraint on $\Omega_{\rm m}$; only an upper limit of 0.09 can be set at
the $1\sigma$ confidence level.

\begin{figure}[hp]
\vskip -0.2in
\centerline{\includegraphics[angle=0,scale=1.2]{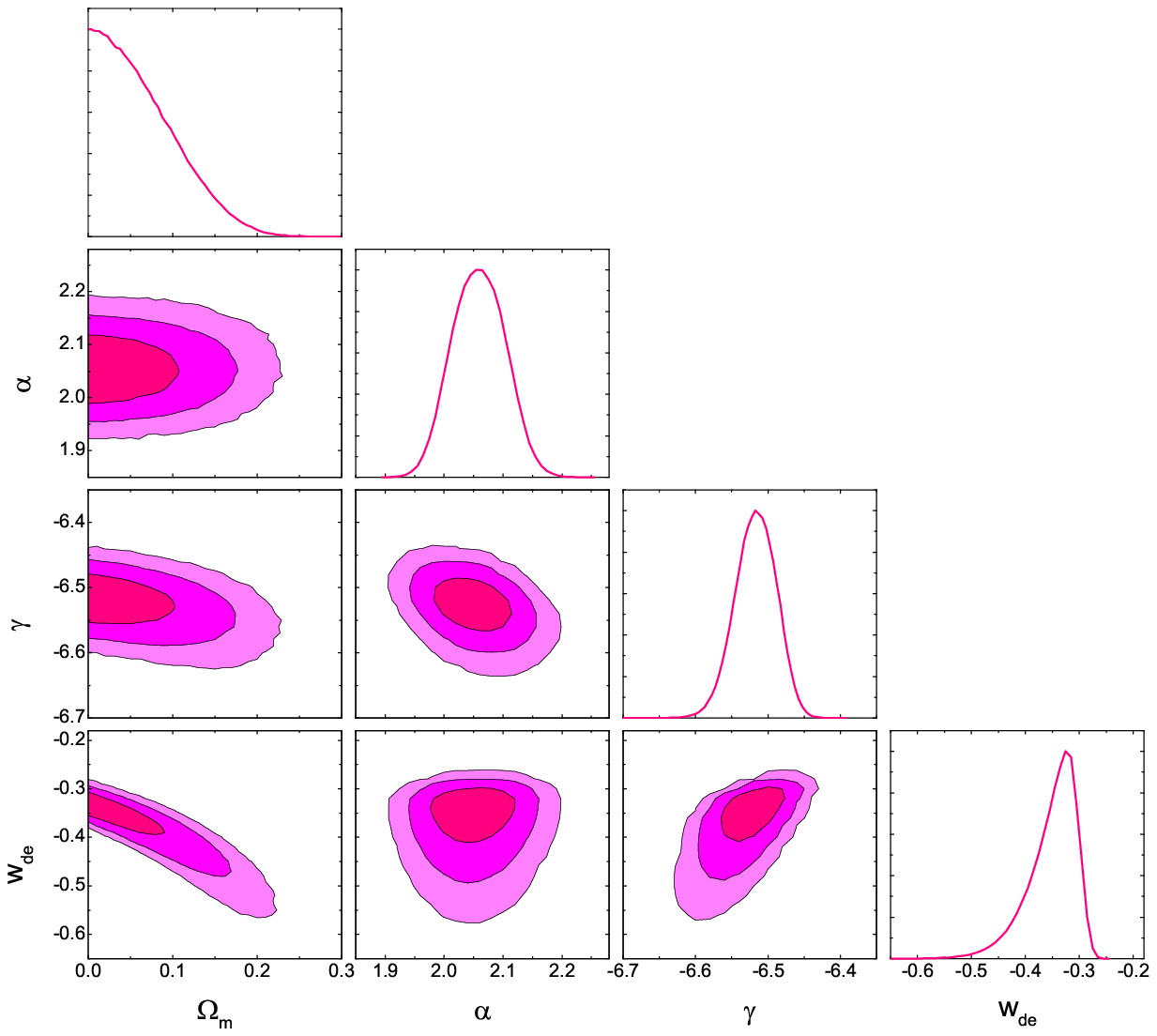}}
\vskip-0.2in
\caption{Same as Figure~5, except now with $R_{\rm h}=ct$ as the (assumed) background
cosmology. The simulated model parameter was $H_{0}=70$ km $\rm s^{-1}$ $\rm Mpc^{-1}$.}
\end{figure}

The corresponding 2-D contours in the $\alpha-\Upsilon$ plane for the $R_{\rm h}=ct$
Universe are shown in Figure~9. The best-fit values for the simulated sample are
$\alpha=2.05_{-0.06}^{+0.07}$ $(1\sigma)$ and $\Upsilon=-6.52_{-0.02}^{+0.02}$ $(1\sigma)$.
These are similar to those in the standard model, but not exactly the same,
reaffirming the importance of reducing the data separately for each model being tested.
With $N=480$, our analysis of the simulated sample shows that in this case the BIC
would favor $R_{\rm h}=ct$ over $w$CDM by an overwhelming likelihood of $99.7\%$
versus only $0.3\%$ (i.e., the prescribed $3\sigma$ confidence limit).

\section{Conclusions}
It is quite evident that SLSNe Ic may be useful cosmological probes, perhaps even
out to redshifts much greater ($z>>2$) than those accessible using Type Ia SNe.
The currently available sample, however, is still quite small; adequate data
to extract correlations between empirical, observable quantities, such as
lightcurve shape, color evolution and peak luminosity, are available only for
tens of events. In this paper, we have proposed to use SLSNe Ic for an
actual one-on-one comparison between competing cosmological models.
This must be done because the results we have presented here already
indicate a strong likelihood of being able to discriminate between models such as
$\Lambda$CDM and $R_{\rm h}=ct$. Such comparisons have already been made
using, e.g., cosmic chronometers (Melia \& Maier 2013), Gamma-ray bursts
(Wei et al. 2013), and Type Ia SNe (Wei et al. 2015).

\begin{figure}[h]
\vskip -0.2in
\centerline{\hskip 1.1in\includegraphics[angle=0,scale=1.0]{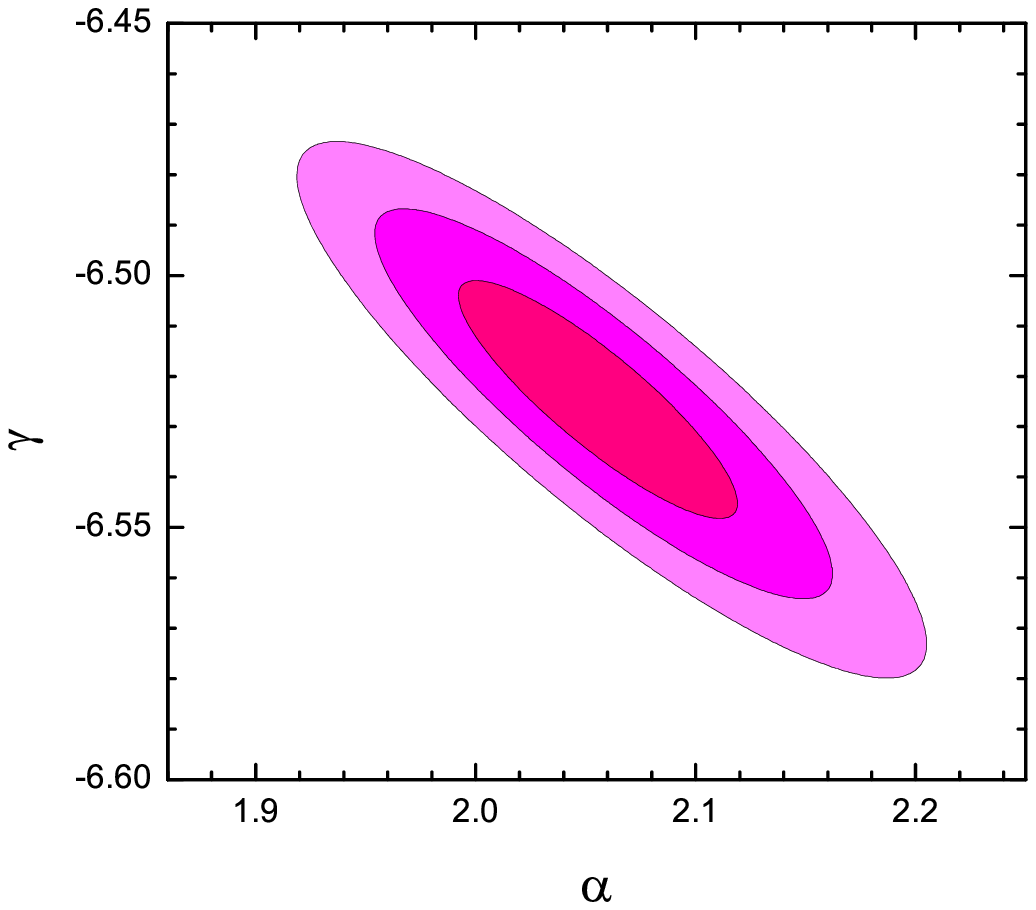}}
\vskip-0.2in
\caption{Same as Figure~7, except now with $R_{\rm h}=ct$ as the (assumed)
background cosmology.}
\end{figure}

We have individually optimized
the parameters in each model by minimizing the $\chi^{2}$ statistic.
With the optimized fits we have reported above, our analysis of the
$\Delta M_{30}$ decline relation (with 11 objects) shows that
$R_{\rm h}=ct$ is favoured over the flat $\Lambda$CDM model with a likelihood of
$\approx 63-74\%$ versus $26-37\%$ (depending on the information criterion).
In our one-on-one comparison using the peak magnitude-color
evolution relation (with 8 objects), the $R_{\rm h}=ct$ Universe
is preferred over $\Lambda$CDM with a likelihood of $\approx 73-82\%$ versus
$18-27\%$.

But though SLSN Ic observations currently tend to favor $R_{\rm h}=ct$
over $\Lambda$CDM, the known sample of such measurements is still too small
for us to completely rule out either model. We have therefore considered
two synthetic samples with characteristics similar to those of the 8 known
SLSN Ic measurements, one based on a $\Lambda$CDM background cosmology, the other on
$R_{\rm h}=ct$. From the analysis of these simulated SLSNe Ic, we have
estimated that a sample of about 240 such events are necessary to rule out
$R_{\rm h}=ct$ at a $\sim 99.7\%$ confidence level if the real cosmology
is in fact $\Lambda$CDM, while a sample of at least 480 SNe would
be needed to similarly rule out $w$CDM if the background cosmology
were instead $R_{\rm h}=ct$. The difference in required sample size
results from $w$CDM's greater flexibility in fitting the data, since
it has a larger number of free parameters.

Our simulations have also shown that a moderate sample size of
$\sim250$ events could reach much tighter constraints on the dark-energy equation
of state $w_{\rm de}$ and on the matter density fraction $\Omega_{\rm m}$
than are currently available with the 580 Union2.1 Type Ia SNe. If SLSNe~Ic
can be commonly detected in the future, they have the potential of greatly
refining the measurement of cosmological parameters, particularly the
dark-energy equation of state $w_{\rm de}$.

Assembling samples of this size may be feasible with upcoming surveys.
For example, the planned survey SUDSS (Survey Using Decam for Super-luminous
Supernovae) using the Dark Energy Camera on the CTIO Blanco 4m telescope
Inserra \& Smartt (2014), and the Subaru/Hyper Suprime-Cam deep survey
(Tanaka et al. 2012), have a goal of discovery several hundred SLSNe out
to $z\sim 4$ over 3 years by imaging tens of square degrees to a limiting
magnitude $\sim 25$ every 2 weeks. These numbers are based on an estimated
production in the local universe (Quimby et al. 2013), who reported a rate of
$32^{+77}_{-26}$ events Gpc$^{-3}$ yr$^{-1}$ at a weighted redshift of
$\langle z\rangle=0.17$. This number is low ($0.01\%$) compared to that
of Type Ia SNe. McCrum et al. (2015) estimate a Type Ic SN rate $\sim 10^{-4}$
of the overall core-collapse SN rate within $0.3<z<1.4$, though this number
could be higher at $z>1.5$ (Cooke et al. 2012) due, perhaps, to a decreasing
metallicity. Moreover,  the increase in cosmic star formation rate would
boost the absolute numbers of SLSNe. As far as observations
from the ground are concerned, assembling a sample of several hundred Type Ic
SLSNe over 3 to 5 years therefore looks quite promising, particulary with
the surveying capability of LSST (see, e.g., Lien \& Fields 2009), which
should cover the whole sky every 2 nights, down to a limiting magnitude
$\sim 24$. The expected rate of discovery of core-collapse SNe with this survey
is expected to be $\sim 1-2$ s$^{-1}$ out to a redshift $z\sim 2$. Given
that $\sim 10^{-4}$ of these are expected to be Type Ic SNe in this
redshift range, one should expect LSST to assemble a $\sim 500$ SLSN
sample in less than a year. The situation from space is even more
exciting. Detecting Type Ic SLSNe out to redshifts $z\sim 10$ via their
restframe 400nm and 520nm bands is plausible with, e.g., EUCLID
(Laureijs et al. 2011), WFIRST and the James Webb Space Telescope
(specifically the NIRCam).\footnote{http://www.stsci.edu/jwst/instruments/nircam/}

A major problem with this approach right now, however, is that one
must rely on the use of a luminosity-limited sample of supernovae, i.e.
those selected to be super-luminous with peak magnitudes $M_{AB}<-21$ mag
and integrated burst energies $\sim$$10^{51}$ erg (Inserra \& Smartt 2014),
in order to test the luminosity distances. The difficulty is that the
luminosity-limited sample may turn out to be different with different
background cosmologies, given that their inferred luminosities are themselves
dependent on the models. This situation is unlikely to change as the sample
grows, so it would be necessary to find a way of identifying these SNe, other
than simply through their magnitudes. Fortunately, in the case of $R_{\rm h}=ct$
versus $\Lambda$CDM, even though their luminosity distances are formulated
differently, it turns out that the ensuing distance measures derived from
these are quite similar all the way out to $z\sim6$. As such, Type Ic SLSNe
that make the cut for one model and not the other are the exception rather
than the rule. In other words, the luminosity-limit applied to the sample
examined here does not bias either model very much. But this may not be
true in general, and an alternative method of selection is highly
desirable.

\acknowledgments We are very grateful to Cosimo Inserra for helpful, clarifying discussions,
and especially to Stephen Smartt for his thoughtful review of the manuscript, which has
led to an improvement in the presentation of our results.
This work is partially supported by the National Basic Research Program (``973" Program) of China
(Grants 2014CB845800 and 2013CB834900), the National Natural Science Foundation of China
(grants Nos. 11322328, 11373068, 11173064 and 11233008), the One-Hundred-Talents Program
and the Youth Innovation Promotion Association, and the Strategic Priority Research Program
``The Emergence of Cosmological Structures'' (Grant No. XDB09000000) of the Chinese Academy
of Sciences, and the Natural Science Foundation of Jiangsu Province (grant no.\ BK2012890).
F.M. is also grateful to Amherst College for its support through a John Woodruff Simpson
Lectureship, and to Purple Mountain Observatory in Nanjing, China, for its hospitality
while this work was being carried out. This work was partially supported by grant
2012T1J0011 from The Chinese Academy of Sciences Visiting Professorships for Senior
International Scientists, and grant GDJ20120491013 from the Chinese State Administration
of Foreign Experts Affairs.


\begin{thebibliography}

\bibitem[]{} Akaike, H. 1973, in Second International Symposium on Information Theory,
eds. B.~N. Petrov and F.~Cs{\'a}ki (Budapest: Akad{\'e}miai Kiad{\'o}), 267

\bibitem[Baltay et al.(2013)]{2013PASP..125..683B} Baltay, C., Rabinowitz,
D., Hadjiyska, E., et al.\ 2013, \pasp, 125, 683

\bibitem[Barbary et al.(2009)]{2009ApJ...690.1358B} Barbary, K., Dawson,
K.~S., Tokita, K., et al.\ 2009, \apj, 690, 1358

\bibitem[Berger et al.(2012)]{2012ApJ...755L..29B} Berger, E., Chornock,
R., Lunnan, R., et al.\ 2012, \apjl, 755, L29

\bibitem[]{} Bhansali, R. J. \& Downham, D. Y. 1977, Biometrika, 64, 547

\bibitem[Cano
\& Jakobsson(2014)]{2014arXiv1409.3570C} Cano, Z., \& Jakobsson, P.\ 2014, arXiv:1409.3570

\bibitem[]{} Cavanaugh, J. E. 1999, Statist. Probab. Lett., 42, 333

\bibitem[Cavanaugh (2004)]{Cav04} Cavanaugh, J. E. 2004, Aust. N.~Z. J. Stat., 46, 257

\bibitem[Chomiuk et al.(2011)]{2011ApJ...743..114C} Chomiuk, L., Chornock,
R., Soderberg, A.~M., et al.\ 2011, \apj, 743, 114

\bibitem[Cooke et al.(2012)]{2012Natur.491..228C} Cooke, J., Sullivan, M.,
Gal-Yam, A., et al.\ 2012, \nat, 491, 228

\bibitem[Drake et al.(2009)]{2009ApJ...696..870D} Drake, A.~J., Djorgovski,
S.~G., Mahabal, A., et al.\ 2009, \apj, 696, 870

\bibitem[Gal-Yam(2012)]{2012Sci...337..927G} Gal-Yam, A.\ 2012, Science,
337, 927

\bibitem[Hamuy et al.(1996)]{1996AJ....112.2438H} Hamuy, M., Phillips,
M.~M., Suntzeff, N.~B., et al.\ 1996, \aj, 112, 2438

\bibitem[Hinshaw et al.(2013)]{2013ApJS..208...19H} Hinshaw, G., Larson,
D., Komatsu, E., et al.\ 2013, \apjs, 208, 19

\bibitem[Hjorth et al.(2003)]{2003Natur.423..847H} Hjorth, J., Sollerman,
J., M{\o}ller, P., et al.\ 2003, \nat, 423, 847

\bibitem[Howell et al.(2013)]{2013ApJ...779...98H} Howell, D. A., Kasen,
D., Lidman, C. et al.\ 2013, ApJ, 779, id.98

\bibitem[Inserra et al.(2013)]{2013ApJ...770..128I} Inserra, C., Smartt,
S.~J., Jerkstrand, A., et al.\ 2013, \apj, 770, 128

\bibitem[Inserra
\& Smartt(2014)]{2014ApJ...796...87I} Inserra, C., \& Smartt, S.~J.\ 2014, \apj, 796, 87

\bibitem[Kaiser et al.(2010)]{2010SPIE.7733E..0EK} Kaiser, N., Burgett, W.,
Chambers, K., et al.\ 2010, \procspie, 7733,

\bibitem[]{} Laureijs, R., Amiaux, J., Arduini, S. et al. 2011, arXiv:1110.3193

\bibitem[Leloudas et
al.(2012)]{2012A&A...541A.129L} Leloudas, G., Chatzopoulos, E., Dilday, B., et al.\ 2012, \aap, 541, A129

\bibitem[Li et al.(2014)]{2014ApJ...796L...4L} Li, X., Hjorth, J.,
\& Wojtak, R.\ 2014, \apjl, 796, LL4

\bibitem[]{} Liddle, A. R. 2004, \mnras, 351, L49

\bibitem[Liddle(2007)]{2007MNRAS.377L..74L} Liddle, A.~R.\ 2007, \mnras,
377, L74

\bibitem[]{} Liddle, A., Mukherjee, P., \& Parkinson, D. 2006, Astron. \& Geophys., 47, 4.30

\bibitem[]{2009JCAP...01..047L} Lien, A. \& Fields, B. D. 2009, \jcap, 1, 047

\bibitem[Lunnan et al.(2013)]{2013ApJ...771...97L} Lunnan, R., Chornock,
R., Berger, E., et al.\ 2013, \apj, 771, 97

\bibitem[McCrum et al.(2015)]{2015MNRAS.448.1206M} McCrum, M., Smartt, S. J.,
Rest, A. et al. 2014, \mnras, 448, 1206

\bibitem[Melia(2007)]{2007MNRAS.382.1917M} Melia, F.\ 2007, \mnras, 382,
1917

\bibitem[Melia(2013)]{2013A&A...553A..76M} Melia, F.\ 2013a, \aap, 553, A76

\bibitem[Melia(2013)]{2013ApJ...764...72M} Melia, F.\ 2013b, \apj, 764, 72

\bibitem[Melia(2014)]{2014JCAP...01..027M} Melia, F.\ 2014, \jcap, 1, 27

\bibitem[Melia(2015)]{2015MNRAS.446.1191M} Melia, F.\ 2015, \mnras, 446, 1191

\bibitem[Melia
\& Abdelqader(2009)]{2009IJMPD..18.1889M} Melia, F., \& Abdelqader, M.\ 2009,
International Journal of Modern Physics D, 18, 1889

\bibitem[Melia
\& Shevchuk(2012)]{2012MNRAS.419.2579M} Melia, F., \& Shevchuk, A.~S.~H.\ 2012, \mnras, 419, 2579

\bibitem[Melia
\& Maier(2013)]{2013MNRAS.432.2669M} Melia, F., \& Maier, R.~S.\ 2013, \mnras, 432, 2669

\bibitem[Nicholl et al.(2013)]{2013Natur.502..346N} Nicholl, M., Smartt,
S.~J., Jerkstrand, A., et al.\ 2013, \nat, 502, 346

\bibitem[Nicholl et al.(2014)]{2014MNRAS.444.2096N} Nicholl, M., Smartt,
S.~J., Jerkstrand, A., et al.\ 2014, \mnras, 444, 2096

\bibitem[Pastorello et al.(2010)]{2010ApJ...724L..16P} Pastorello, A.,
Smartt, S.~J., Botticella, M.~T., et al.\ 2010, \apjl, 724, L16

\bibitem[Perlmutter et al.(1998)]{1998Natur.391...51P} Perlmutter, S.,
Aldering, G., della Valle, M., et al.\ 1998, \nat, 391, 51

\bibitem[Perlmutter et al.(1999)]{1999ApJ...517..565P} Perlmutter, S.,
Aldering, G., Goldhaber, G., et al.\ 1999, \apj, 517, 565

\bibitem[Perlmutter et al.(1997)]{1997ApJ...483..565P} Perlmutter, S.,
Gabi, S., Goldhaber, G., et al.\ 1997, \apj, 483, 565

\bibitem[Phillips(1993)]{1993ApJ...413L.105P} Phillips, M.~M.\ 1993, \apjl,
413, L105

\bibitem[Pskovskii(1977)]{1977SvA....21..675P} Pskovskii, I.~P.\ 1977,
\sovast, 21, 675

\bibitem[]{Quimby2005} Quimby, R. M., Castro, F., Gerardy, C. L. et al. 2005,
BAAS, 37, \#171.02

\bibitem[Quimby(2006)]{2006CBET..644....1Q} Quimby, R.\ 2006, Central
Bureau Electronic Telegrams, 644, 1

\bibitem[Quimby et al.(2011)]{2011Natur.474..487Q} Quimby, R.~M., Kulkarni,
S.~R., Kasliwal, M.~M., et al.\ 2011, \nat, 474, 487

\bibitem[Quimby et al.(2007)]{2007ApJ...668L..99Q} Quimby, R.~M., Aldering,
G., Wheeler, J.~C., et al.\ 2007, \apjl, 668, L99

\bibitem[Quimby et al.(2013)]{2013ApJ...768L..20Q} Quimby, R. M., Werner,
M. C., Oguri, M. et al. 2013, \apjl, 768, LL20

\bibitem[Rau et al.(2009)]{2009PASP..121.1334R} Rau, A., Kulkarni, S.~R.,
Law, N.~M., et al.\ 2009, \pasp, 121, 1334

\bibitem[Riess et al.(1998)]{1998AJ....116.1009R} Riess, A.~G., Filippenko,
A.~V., Challis, P., et al.\ 1998, \aj, 116, 1009

\bibitem[Rust(1974)]{1974BAAS....6..309R} Rust, B.~W.\ 1974, \baas, 6, 309

\bibitem[Samushia
\& Ratra(2009)]{2009ApJ...703.1904S} Samushia, L., \& Ratra, B.\ 2009, \apj, 703, 1904

\bibitem[Schmidt et al.(1998)]{1998ApJ...507...46S} Schmidt, B.~P.,
Suntzeff, N.~B., Phillips, M.~M., et al.\ 1998, \apj, 507, 46

\bibitem[Schwarz (1978)]{Sch78} Schwarz, G. 1978, Ann. Statist., 6, 461

\bibitem[Shao et al.(2011)]{2011ApJ...738...19S} Shao, L., Dai, Z.-G., Fan,
Y.-Z., et al.\ 2011, \apj, 738, 19

\bibitem[Smith et al.(2007)]{2007ApJ...666.1116S} Smith, N., Li, W., Foley,
R.~J., et al.\ 2007, \apj, 666, 1116

\bibitem[Stanek et al.(2003)]{2003ApJ...591L..17S} Stanek, K.~Z., Matheson,
T., Garnavich, P.~M., et al.\ 2003, \apjl, 591, L17

\bibitem[Suzuki et al.(2012)]{2012ApJ...746...85S} Suzuki, N., Rubin, D.,
Lidman, C., et al.\ 2012, \apj, 746, 85

\bibitem[]{} Tan, M. Y. J. \& Biswas, R. 2012, MNRAS, 419, 3292

\bibitem[]{} Tanaka, M., Moriya, T. J., Yoshida, N. \& Nomoto, K. 2012,
MNRAS, 422, 2675

\bibitem[]{} Takeuchi, T. T. 2000, Ap\&SS, 271, 213

\bibitem[Wei et al.(2013)]{2013ApJ...772...43W} Wei, J.-J., Wu, X.-F.,
\& Melia, F.\ 2013, \apj, 772, 43

\bibitem[Wei et al.(2015)]{2015AJ....149..102W} Wei, J.-J., Wu, X.-F.,
Melia, F. \& Maier, R. S.\ 2015, \aj, 149, 102

\end{thebibliography}
\end{document}